\documentstyle[aps,tighten,graphicx,prl,twocolumn]{revtex}

\begin{document}
\title{  
Inseparability criteria for demonstration of the Einstein-Podolsky-Rosen 
gedanken experiment\\ }
 \vskip 1 truecm
\author{M. D. Reid\\ }
\address{Physics Department, University of Queensland, Brisbane, Australia\\ }  	  
\date{\today}
\maketitle
\vskip 1 truecm
\begin{abstract}
It is shown that a criterion used to demonstrate realization of the 1935 
 Einstein-Podolsky-Rosen (EPR) gedanken experiment  
is sufficient 
to demonstrate quantum entanglement.
 A further set of measurable criteria sufficient to demonstrate  
EPR gedanken experiment is proposed, these being the set of criteria 
sufficient to 
demonstrate entanglement, by way of a measured violation of a necessary 
condition of separability. In this way, provided the
 spatial separation of systems is sufficient to ensure EPR's  
locality hypothesis, it is shown how a measured demonstration of 
entanglement will, at least, be equivalent to a demonstration of the EPR 
gedanken experiment. Using hidden variables it is explained how such 
demonstrations are a direct 
manifestation of the inconsistency of local realism with quantum 
mechanics. 

\end{abstract}
\narrowtext
\vskip 0.5 truecm

In 1935 Einstein, Podolsky and 
Rosen$^{\cite{1}}$ (EPR) defined a premise called local realism. 
They showed that,
 for certain correlated 
spatially separated systems, if quantum mechanics is to be consistent
 with local 
realism,
the position and momentum of a 
single particle must be simultaneously defined to a precision beyond 
the bounds given by 
 the uncertainty principle. EPR took the view 
 that local realism must be valid and therefore argued that      
 quantum mechanics was incomplete.  
 
 Schrodinger's reply$^{\cite{2}}$ in 1935 is now also well-known. 
 In this reply 
 Schrodinger introduced the concept of entangled quantum states, 
 referring to their paradoxical nature for separated systems. Quantum 
 entanglement is now a concept fundamental to all aspects of quantum information 
 theory. 
 
  A criterion to demonstrate EPR correlations$^{\cite{3}}$ for real 
  experiments was 
 presented in 1989. 
 The first experimental achievement$^{\cite{4}}$ of 
 this EPR criterion, for 
   measurements with 
   continuous variable outcomes not impeded by detection inefficiencies,
    was presented by Ou et al  
 in 1992. Here the two conjugate quantities are the 
 quadrature phase amplitudes of the field$^{\cite{5}}$, represented quantum 
 mechanically by a quantized harmonic oscillator. 
There have now been further experimental observations$^{\cite{6,7}}$ of 
continuous variable EPR correlations, and also proposals for 
further experiments$^{\cite{8}}$.
While these EPR fields
  have proven significant in enabling the experimental 
  realization of
   continuous variable quantum teleportation$^{\cite{9}}$, 
 and may have application also to quantum cryptography$^{\cite{10}}$, 
 the EPR experiments are also significant in providing a conclusive 
 demonstration of the inconsistency of local realism with quantum 
 mechanics, as we elaborate on in this paper.

  My objective is to formalize the link between the original 1935 
  EPR argument and Schrodinger's 1935 essay introducing
  entanglement, by showing how the experimental demonstration 
  of these EPR correlations will 
  correspond to a
   demonstration of entanglement, and vice versa, provided 
   measurements and spatial 
   separations  are sufficient to ensure the EPR locality hypothesis. 
   This provides a means to demonstrate objectively
    the inconsistency of quantum entangled states 
   with local realism.
   
   In order to link EPR correlations with entanglement, I first 
    consider a 
   generalized EPR 
   argument applying to situations of less than 
   maximum correlation. 
 It is then proven that the 
   1989 EPR criterion will always imply entanglement. It follows that continuous 
 variable entanglement, an unconditional entanglement defined in the 
 sense of Turchette et al$^{\cite{11}}$, has been experimentally 
 observed$^{\cite{4,6,7,12}}$. 
 
 I next  
show that the local realistic
 description of a violation of certain sets of  
constraints, these being a set of  necessary criteria for 
separability$^{\cite{13}}$, 
 necessitates a 
decomposition into local substates that, individually, defy the 
quantum bound set  by the 
uncertainty relation. This is precisely the criterion for 
realization of an EPR gedanken experiment.
It also follows from this logic that the entire quantum state 
predicting these violations must be inseparable, and in this way, with 
some further generalizations, we arrive at  
 our result.

 Crucial to demonstrating EPR correlations is the definition of local 
 realism$^{\cite{1}}$. 
 The premise of realism implies 
 that if one can predict with 
 certainty the result of a measurement of a physical quantity 
 at a location $A$, without disturbing the 
 system at $A$, then the results of the measurement were 
 predetermined. There is an 
 ``element of reality'' corresponding to this 
 physical quantity, the  element of reality being a variable that assumes 
 one of the set of values that are the predicted results of the 
 measurement.  The locality assumption postulates that measurements 
at a spatially separated location $B$ cannot immediately influence the 
subsystem at $A$. 
 
 EPR argued as follows. Consider two 
 observables $\hat{x}$ and $\hat{p}$ for subsystem $A$, where $\hat{x}$ 
 and $\hat{p}$ satisfy an 
 uncertainty relation $\Delta \hat{x} \Delta \hat{p}\geq C $.
 Suppose one may predict with certainty the result of 
 measurement $\hat{x}$ by a measurement performed at $B$, 
 or alternatively,
  for a different measurement at $B$, the 
 result of measurement  $\hat{p}$. Assuming ``local realism'',
  we 
 deduce the existence 
 of an element of reality, ${\tilde x}$, for 
 the physical quantity $\hat{x}$; 
 and also 
 an element of 
 reality, ${\tilde p}$, for $\hat{p}$. 
   Local realism then implies the 
  existence of two hidden variables ${\tilde x}$ and ${\tilde p}$ that 
  simultaneously 
  determine, with no uncertainty, the values for the result of an 
  $\hat{x}$ or $\hat{p}$ measurement on subsystem $A$, should it be performed. 
  This 
  hidden variable state for the subsystem $A$ is not describable within 
  quantum mechanics, because of the 
    uncertainty relation.

For a link with inseparability, we extend the 
EPR argument to situations where the result of measurement $\hat{x}$ at $A$ 
cannot be predicted with absolute certainty$^{\cite{3}}$. 
 Local 
  realism still allows us to deduce the existence of an element 
  of reality (${\tilde x}$) for $\hat{x}$ at $A$,
   since we can make a prediction of the 
  result at $A$, without disturbing the subsystem at $A$, under the locality 
   assumption. This prediction is based on the result $y_{i}$ of a 
   measurement $\hat{y}$ performed 
   at $B$.
  The possible values for the ``element 
  of reality'', the predicted results of the measurement $\hat{x}$, 
  are no longer a set of definite 
  numbers with zero uncertainty, but are a set 
  of distributions, one for each 
  possible result $y_{i}¥$ for $\hat{y}$ at $B$. 
      The element of reality $\tilde{x}$ 
   becomes indeterminate, having a finite variance.

    Each $y_{i}$ represents a possible 
  hidden variable state for the subsystem $A$.
  The predicted probability of a result for the measurement 
 $\hat{x}$ at $A$, should the subsystem $A$ be in the hidden 
 variable state $y_{i}$, is given by the distribution labeled by $y_{i}$. 
  This  
 distribution is  given formally by 
 the conditional probability $P(x/y_{i})$, the probability of obtaining a 
 result $x$ upon measurement of $\hat{x}$ at $A$, given the result 
 $y_{i}¥$ for 
 the measurement of $\hat{y}$ at $B$. 
  The  
 probability that the subsystem $A$ is in the hidden variable state designated 
 $y_{i}$ is $P(y_{i})$, the probability of 
 the result $y_{i}¥$ at $B$, since through locality the 
 action of measuring $\hat{y}$ at $B$ could not have 
 induced the result at $A$. 
  We attribute to the inferred element of reality ${\tilde x}$, based 
  on measurements $\hat{y}$, the  
 weighted variance 
 $\Delta_{inf,min}^{2}\hat{x}=\sum_{y_{i}¥}P(y_{i})\Delta_{i}¥^{2}¥$ 
 where $\mu_{i}¥$ and 
 $\Delta_{i}$  are 
 the mean and standard deviation, respectively, of the conditional distribution 
 $P(x/y_{i})$.
 
 The   best estimate$^{\cite{14}}$ of the outcome of 
 $\hat{x}$ at $A$, based on a result $y_{i}¥$ for the   
 measurement at $B$, is given by $\mu_{i}$,  
 and $\Delta_{i}=\sqrt{\langle (x/y_{i}-\mu_{i})^{2}\rangle}$ 
 is the
 root mean square of the error or deviation 
 $\delta_{i}=x/y_{i}-\mu_{i}¥$  
 in the inference $\mu_{i}$. (Here $x/y_{i}$ is the value obtained for $x$ 
 given the result $y_{i}$ at $B$). 
  $\Delta_{inf,min}^{2}\hat{x}$ defines the (minimum) average variance 
  for   
  the inference of the result of a 
 measurement $\hat{x}$ at $A$, based on the result of the measurement 
 $\hat{y}$ at $B$.
  Similarly for the inference of the result of  
 measurement $\hat{p}$, based on a different measurement at $B$, 
 we define a $\Delta_{inf,min}\hat{p}$. 
 
We consider a measured error 
 $\Delta_{inf}\hat{x}$ in the prediction for the outcome of measurement $\hat{x}$ 
 at $A$, based on a result at $B$; and a similar measured error 
 $\Delta_{inf}\hat{p}$ for the prediction of $\hat{p}$ at $A$. 
 The 1989    criterion for demonstration of EPR correlations is to find
\begin{equation}  
\Delta_{inf}\hat{x}\Delta_{inf}\hat{p}<C
\end{equation} 
since here elements of reality ${\tilde x}$, ${\tilde p}$ 
simultaneously attributed to 
system $A$ by local realism are incompatible with the uncertainty principle.

 We now show that the EPR 
criterion (1) is sufficient to demonstrate quantum
 entanglement.  To do this we show that a separable quantum state,  
 defined as expressible by a 
 density matrix of the form
     \begin{equation}
     \rho= \sum_{r}P_{r}\rho_{r}^{A}\rho_{r}^{B}
\end{equation}
 where $\sum_{r} P_{r}=1$,
will imply $\Delta_{inf}\hat{x}\Delta_{inf}\hat{p}\geq C$.
 The  
conditional probability of 
result $x$ for measurement $\hat{x}$ at $A$ given a simultaneous 
measurement of $\hat{y}$ at $B$ with result $y_{i}$ is   
$P(x/y_{i}¥)=P(x,y_{i}¥)/P(y_{i}¥)$ where, given (2) 
     \begin{eqnarray}
     P(x,y_{i}¥)
    &=&\sum_{r}¥P_{r}¥P_{r}(y_{i})P_{r}(x)
 \end{eqnarray}
Here $|x\rangle,|y\rangle$ are the eigenstates of $\hat{x}$,$\hat{y}$ 
respectively, and $P_{r}(x)=\langle x|\rho_{r}^{A}|x\rangle$, 
$P_{r}¥(y_{i}¥)=\langle y_{i}¥|\rho_{r}^{B}|y_{i}¥\rangle$.
The mean $\mu_{i}¥$ of this conditional distribution is  
     $\mu_{i}=\sum_{x}xP(x/y_{i})
     =\{\sum_{r}P_{r}P_{r}¥(y_{i}) \langle x\rangle_{r}\}/P(y_{i})
$ where $\langle x\rangle_{r}=\sum_{x}xP_{r}¥(x)$.
  The variance $\Delta_{i}^{2}¥$ of the distribution $P(x/y_{i}¥)$ is 
$\Delta_{i}^{2}
       =\{\sum_{r}¥P_{r}¥P_{r}(y_{i})\sum_{x}P_{r}¥(x)
      (x-\mu_{i}¥)^{2}\}/P(y_{i})$. 
      For each state $r$, the mean square 
deviation $\sum_{x}P_{r}(x)(x-d)^{2}$ 
is minimized with the choice 
$d=\langle x\rangle_{r}$$^{\cite{14}}$.  
Therefore for the choice $d=\mu_{i}¥$,
$     \Delta_{i}^{2}\geq
    \{\sum_{r}¥P_{r}¥P_{r}(y_{i})\sum_{x}
     P_{r}¥(x)(x-\langle x\rangle_{r})^{2}\}/P(y_{i})
     =\{\sum_{r}¥P_{r}¥P_{r}¥(y_{i}¥)\sigma_{r}^{2}¥(x)\}/P(y_{i})
$ where $\sigma_{r}^{2}¥(x)$ is the variance of $P_{r}(x)$.
 Taking the average variance over the 
$y_{i}$ we get 
\begin{eqnarray}
\Delta_{inf}^{2}¥\hat{x}
&\geq&\sum_{y_{i}}P(y_{i})\{\sum_{r}P_{r}¥P_{r}(y_{i})
                \sigma_{r}^{2}(x)\}/P(y_{i})\nonumber\\
&=&\sum_{r}P_{r}\sigma_{r}^{2}¥(x)\sum_{y_{i}¥}P_{r}(y_{i})\nonumber\\
&=&\sum_{r}P_{r}\sigma_{r}^{2}¥(x)
\end{eqnarray}
 Also $\Delta_{inf}^{2}\hat{p}\geq\sum_{r} P_{r}\sigma_{r}^{2}¥(p)$, where 
 $\sigma_{r}^{2}(p)$ is the variance of $P_{r}(p)=\langle 
 p|\rho_{r}^{A}|p\rangle$, $|p\rangle$ being the eigenstate of 
 $\hat{p}$. This implies (from the Cauchy-Schwarz inequality)  
  $\Delta_{inf}^{2}\hat{x}\Delta_{inf}^{2}\hat{p}\geq 
  \{\sum_{r}P_{r}\sigma_{r}^{2}¥(x)\}  \{\sum_{r}P_{r}\sigma_{r}^{2}¥(p)\}
  \geq| \sum_{r}P_{r}\sigma_{r}¥(x)
   \sigma_{r}(p) |^{2}$. 
 For any $\rho_{r}^{A}$ it is constrained, by the uncertainty relation, that  
$  \sigma_{r}(x)\sigma_{r}(p)\geq C$. We conclude that for a 
separable quantum state
 \begin{equation}
  \Delta_{inf}\hat{x}\Delta_{inf}\hat{p}\geq C. 
\end{equation}

  The evaluation of the conditional probability 
  distribution  
 for each outcome $y_{i}$ at $B$ is not always be practical. 
 We might 
  propose the linear estimate  
 $x_{est}=gy_{i}+d$ ($g$ and $d$ are constants) for the result $x$ 
 at $A$, 
 given a 
 result $y_{i}¥$ for the measurement at $B$. 
 The size of the 
deviation $\delta¥=x-(gy_{i}+d)$ can 
 be measured. We simultaneously measure $\hat{x}$ at $A$ and 
$\hat{y}$ at $B$, to determine $x$ and $y_{i}$ and then 
calculate for a given $y_{i}$,  
$\langle\delta^{2}\rangle_{i}¥=\sum_{x}¥ 
P(x/y_{i})\{x-(gy_{i}+d)\}^{2}$. 
Averaging over the different values of $y_{i}$ we obtain as a measure 
of error in our inference, based on the linear 
estimate: 
$\Delta_{inf,L}^{2}\hat{x}=\sum_{y_{i}¥}P(y_{i})
\langle\delta^{2}\rangle_{i}=
\sum_{x,y_{i}¥}P(x,y_{i}¥)\{x-(gy_{i}+d)\}^{2}
=\langle\{\hat{x}-(g\hat{y}+d)\}^{2}\rangle
$. The best linear estimate $x_{est}$ is the one that will minimize  
$\Delta^{2}_{inf,L}\hat{x}$.  This corresponds to the choice$^{\cite{14}}$ 
$d=-\langle (\hat{x}-g\hat{y})\rangle$. (Denoting 
$\delta_{0}=\hat{x}-g\hat{y}¥$, our choice of estimate optimized with 
respect to $d$ gives a minimum error 
$\Delta^{2}_{inf,L}\hat{x}=
\langle \delta^{2}_{0}¥-\langle\delta_{0}¥\rangle^{2}¥\rangle$.) 
The best choice for $g$ is discussed in [3]. The 
quantity $\Delta_{inf,L}^{2}\hat{x}$ may be measured 
straightforwardly, as discussed in [3] and [4,6,7].

If the estimate $x_{est}¥$ corresponds to the mean of the 
conditional distribution 
$P(x/y_{j}¥)$ then 
the variance $\Delta^{2}_{inf,L}\hat{x}$ will correspond to 
the average conditional variance $\sum_{y_{i}¥}P(y_{i})\Delta_{i}¥^{2}¥$ 
specified above. This is the case, with a certain choice of $g$, for the two-mode 
squeezed state used to model continuous variable 
EPR states generated to date.
In general the 
variances of type $\Delta^{2}_{inf,L}\hat{x}$ 
based on estimates will be greater than or equal to the optimal
 evaluated from the conditionals, and the 
  separable quantum state 
must predict  $ \Delta_{inf,L}\hat{x}\Delta_{inf,L}\hat{p}\geq C$.
       The EPR criterion (1) so measured then implies inseparability,
        for any $g$ and $d$. 
      To show explicitly (optimizing $d$, 
       but keep $g$ general), separability implies (use [14])   
      \begin{eqnarray}
  &\quad&    \Delta^{2}_{inf,L}\hat{x}\geq\langle
       \{\hat{x}-\langle \hat{x}\rangle-g(\hat{y}-\langle 
       \hat{y}\rangle)\}^{2}
       \rangle\nonumber\\
       &=&\sum_{x,y}\sum_{r}¥P_{r}¥\langle x|\langle 
      y|\rho_{r}^{A}¥\rho_{r}^{B}¥
      \{\hat{x}-\langle\hat{x}\rangle-g(\hat{y}-\langle \hat{y}\rangle)\}^{2}
      |x\rangle |y\rangle \nonumber\\
         &=&\sum_{r}¥P_{r}¥\langle(\hat{\delta_{0}¥}-\langle 
      \hat{\delta_{0}}\rangle)^{2}
     \rangle_{r}
     \geq \sum_{r}P_{r}
     \langle(\hat{\delta_{0}}-\langle \hat{\delta_{0}}\rangle_{r}¥)^{2}
     \rangle_{r}
 \end{eqnarray}
Here $\hat{\delta}_{0}¥=\hat{x}-g\hat{y}$ and $\langle 
q\rangle_{r}$ denotes the average  for state $r$ given by density 
operator $\rho_{r}=\rho_{r}^{A}\rho_{r}^{B}$. Since  $\rho_{r}$ 
factorizes, 
$\langle\hat{x}\hat{y}\rangle_{r}¥=\langle \hat{x} 
\rangle_{r}¥\langle\hat{y}\rangle_{r}¥$.  
We have $
      \Delta^{2}_{inf,L}\hat{x}\geq \sum_{r}P_{r}(\langle 
      \hat{\delta^{2}_{0}¥}\rangle_{r}¥-\langle\hat{\delta_{0}¥}\rangle_{r}^{2}¥)
      = \sum_{r}P_{r}(\Delta^{2}_{r}\hat{x}+g^{2}¥
     \Delta^{2}_{r}\hat{y})$
where $\Delta^{2}_{r}\hat{x}=\sigma_{r}^{2}¥(x)$ and 
$\Delta^{2}_{r}\hat{y}=\langle 
\hat{y}^{2}\rangle_{r}-\langle\hat{y}\rangle_{r}^{2}$.
Also  
$      \Delta^{2}_{inf,L}\hat{p}
     \geq\sum_{r}P_{r}(\Delta^{2}_{r}\hat{p}+h^{2}
     \Delta^{2}_{r}\hat{q})$ where 
     $\Delta^{2}_{r}\hat{p}=\sigma_{r}^{2}(p)$ and $\hat{q}$ is the 
     measurement at $B$ used to infer the result for 
     $\hat{p}$ at $A$.
 It follows (take $\Delta \hat{y} \Delta \hat{q}\geq D $)
$ \Delta^{2}_{inf,L}\hat{x}\Delta^{2}_{inf,L}\hat{p}
       \geq\sum_{r}P_{r}¥\{ 
      \sigma_{r}^{2}(x)+g^{2}\Delta^{2}_{r}\hat{y}\} 
      \sum_{r}P_{r}¥ 
      \{\sigma_{r}^{2}(p)+h^{2}\Delta^{2}_{r}\hat{q}\}$. Separability 
      implies 
      \begin{eqnarray}
      \Delta^{2}_{inf,L}\hat{x}\Delta^{2}_{inf,L}\hat{p}
&\geq& (C^{2}+g^{2}h^{2}D^{2})
\end{eqnarray}
 
 I now propose a general method to demonstrate the EPR gedanken 
 experiment, by which the incompatibility of the  
  local realistic elements of reality with
  the uncertainty relation can be {\it inferred}, by way of violations of 
  certain sets of constraints. We {\it propose}, 
   without measurement, the 
   existence of elements of reality, 
   leaving {\it unspecified} the values for 
   the variances of  
   the elements of reality.  
    At $A$ one measures either $x$ 
or $p$, a choice denoted by different values, $0$ and $\pi$ 
respectively, of a 
parameter $\theta$. At $B$, simultaneously, there is the  
choice, denoted by $\phi$,
 to measure an $x_{B}$ or 
$p_{B}$.  
 The set of 
 elements of reality ${\tilde x},{\tilde p},\ldots$ for subsystem $A$, and 
  ${\tilde x}_{B},{\tilde p}_{B}¥,\ldots$ for 
  subsystem $B$,
 form a set of hidden variables $\{\lambda\}$ for the entire system, with a 
probability distribution $\rho(\lambda)$. 
For each hidden variable state, there is a probability 
$p_{x}^A(\theta, \lambda )$ 
(independent of 
$\phi$ and with unspecified variance) 
for the result $x$ of 
measurement $\theta$ at $A$. Similarly a $p_{y}^B(\phi, \lambda )$ is defined.

Assuming a general local 
hidden variable theory then, either as a consequence of EPR's local realism 
or as a new more general definition, the joint probability 
$P_{\theta,\phi}(x,y)$ of obtaining an outcome $x$ 
at $A$ and $y$ at $B$ is 
\begin{eqnarray}
P_{\theta ,\phi}(x,y)= \int_{\lambda}¥ \rho(\lambda) \quad 
p_{x}^A(\theta, \lambda ) 
p_{y}^B(\phi, \lambda )\quad d\lambda  
\end{eqnarray}
  Such local hidden variable theories were 
  considered by Bell$^{\cite{15}}$. We also propose an auxiliary 
  assumption, to now specify the variances of the elements of 
  reality, by proposing that for each
 hidden variable state $\{\lambda\}$, the variances 
$\sigma^{2}_{\lambda}(x)$, $\sigma^{2}_{\lambda}(p)$ of
 $p_{x}^{A}(\theta=0,\lambda)$, $p_{x}^{A}¥(\theta=\pi,\lambda)$
respectively are restricted by the quantum ``uncertainty principle'' 
bound
\begin{equation} 
\sigma_{\lambda}(x)\sigma_{\lambda}(p)\geq C
\end{equation}
  Following the logic of (3) to (5), the local realistic theory (8) 
  with assumption (9) will imply
 $\Delta_{inf}\hat{x} 
  \Delta_{inf}\hat{p}\geq C$.
  In fact the general local realistic theory assumption 
  (8), being separable in form, 
  with the proviso (9) (and alternative provisos restricting the possible 
  outcomes for a given hidden variable state 
  to be within the domain predicted by a 
  quantum state), will predict a whole set of 
  inequalities or 
  constraints, these being precisely the set of criteria derivable 
  from the assumption of general quantum separability (2). For example from 
  (7) above, quantum separability (2) implies (for any $g$) 
  $\langle
       \{\hat{x}-\langle \hat{x}\rangle-g(\hat{y}-\langle 
       \hat{y}\rangle)\}^{2}¥
       \rangle\langle
       \{\hat{p}-\langle \hat{p}\rangle-g(\hat{q}-\langle 
       \hat{q}\rangle)\}^{2}¥
       \rangle \geq C^{2}(1+g^{4})$, and this also follows from 
      assumptions (8) and (9), as do the  necessary conditions following from 
      quantum separability where the uncertainty bound is used, derived 
      in recent work by Duan et al$^{\cite{13}}$, and 
      Simon$^{\cite{13}}$.
    
  The demonstration of entanglement may be defined as the measured 
  experimental violation of any one of the  
  set of necessary criteria for separability (the results derivable from 
  the general separable form (2)).
   Provided measurements and  
   spatial separations between subsystems $A$ and $B$ allow   
  justification of the locality assumption, 
  such violations rule out, at least, the validity of all local 
  hidden variables theories (8) where the hidden variable states 
  (elements of reality) satisfy for each subsystem at $A$, 
  and $B$, the bound (9) given by the uncertainty relation (or an 
  alternative quantum bound). 
  These violations 
  are then none other than a demonstration of an
  EPR gedanken experiment, and lead to 
  EPR's conclusion: that the predictions of 
  quantum mechanics can only be represented by local realism, if the 
  localized systems at $A$ ($B$) that are necessarily part of the 
  local realistic theory are described by something other 
  than a quantum state satisfying the quantum bound (9). 
     Quantum mechanics can only be a 
  local realistic theory, if it is 
   ``completed'' to allow a violation of (9). In this way the inconsistency of 
   quantum mechanics with local realism, a demonstration of the EPR 
   paradox, is objectively demonstrable through  
   entanglement criteria. 
   
These more general entanglement criteria for demonstrating the EPR 
gedanken experiment  
are useful, since for example by (7) for $g=1$ one 
needs only prove a two-mode squeezing result 
(of type  $\Delta^{2}_{inf,L}\hat{x}
< 2C^{2}$), averaged for both conjugate 
operators, to obtain        $ \Delta_{inf,L}\hat{x}\Delta_{inf,L}\hat{p}
< 2C^{2}$; as opposed to searching for a $ \Delta_{inf}\hat{x}\Delta{inf}\hat{p}
< C^{2}$ with an optimal $g$.

  Since the separable quantum state (2) 
  satisfies the local hidden variable decomposition  
  (8) with (9) satisfied (``completed'' quantum states do not actually 
  exist in quantum theory), it also follows from the very nature of the
   EPR argument that its 
  demonstration can only come from quantum states that are 
  inseparable. 
  
  A subset, these being the Bell-type inequalities$^{\cite{15}}$, of the necessary 
   criteria following from the 
  local hidden variable decomposition (8) 
  (and also from quantum separability (2)) do not require the 
  additional assumption of a quantum bound (9). The conclusions 
  drawn from such demonstrations of entanglement are 
  stronger, in that all local realistic
   theories (8) are ruled out, even those ``completing'' quantum 
   mechanics. Local realism itself is proved 
  incorrect.

It has been shown possible in some cases
 to predict EPR correlations satisfying (1) 
from a local hidden variable 
theory, derived from the quantum Wigner function, that gives agreement 
with the quantum predictions for the direct $x$, $p$ measurements. 
This implies that certain  Bell inequalities for $\hat{x},\hat{p}$ 
measurements will not be violated in this 
case. 
This might lead to the interpretation that the EPR experiment 
itself, demonstrating (1),  
reflected a situation in which quantum and local realistic (classical)
 domains are not 
distinguishable. This is not the case.  The local realistic 
hidden variable theory 
used to give the quantum predictions    
is, necessarily, not actually quantum theory, since it  must 
incorporate a description $\{\lambda_{a}\}$ 
for a state of the system at $A$ or $B$ 
in which the $x$ and $p$ are prespecified to a variance better than 
the uncertainty principle. The separable local hidden variable theory 
based on the Wigner function is not a separable (local) theory in 
quantum mechanics since these simultaneously well-defined $x$ and 
$p$ are not quantum states.

 This is generally true where we have experimental 
 violations of the necessary criteria for 
 separability, whether viewed as a demonstration of entanglement or of 
 EPR correlations; that it is proved that local realism can only be retained 
 through certain theories alternative to quantum  mechanics. In this way, 
 the inconsistency of quantum 
 mechanics with local realism is demonstrated through entanglement, 
 it being the purpose of the subset 
 Bell-type tests based only on (8) to rule out these further 
 alternatives.
 
 I am grateful to P. 
 D. Drummond and H. J. Kimble for stimulating comments.


\begin{references}
\bibitem{14} This choice $d=\langle x \rangle$ will minimize  
  $V=\langle (x-d)^{2}\rangle$ and therefore the rms error (take 
  $\partial V/\partial d$).
\end{references}
\end{document}